\documentclass[pdflatex,
            sn-mathphys-num,
            iicol]{sn-jnl}
\usepackage{graphicx}%
\usepackage{multirow}%
\usepackage{amsmath,amssymb,amsfonts}%
\usepackage{amsthm}%
\usepackage{mathrsfs}%
\usepackage[title]{appendix}%
\usepackage{xcolor}%
\usepackage{textcomp}%
\usepackage{manyfoot}%
\usepackage{booktabs}%
\usepackage{algorithm}%
\usepackage{algorithmicx}%
\usepackage{algpseudocode}%
\usepackage{listings}%
\usepackage{tikz}
\usepackage{subcaption}
\usepackage{overpic}
\usepackage{siunitx}
\usepackage[sort=none]{glossaries-extra}
\setabbreviationstyle[nameonly]{short-nolong}
\glssetcategoryattribute{nameonly}{nohyper}{true}
\glssetcategoryattribute{common}{nohyper}{true}

\newabbreviation[category=common]{DM}{DM}{Dark Matter}
\newabbreviation[category=common]{CDM}{CDM}{Cold Dark Matter}
\newabbreviation[category=common]{HDM}{HDM}{Hot Dark Matter}
\newabbreviation[category=common]{WDM}{WDM}{Warm Dark Matter}
\newabbreviation[category=common]{CMB}{CMB}{Cosmic Microwave Background}
\newabbreviation[category=common]{SM}{SM}{Standard Model}
\newabbreviation[category=common]{MACHOS}{MACHOs}{massive compact halo objects}
\newabbreviation[category=common]{LY}{LY}{light yield}
\newabbreviation[category=common]{WIMPs}{WIMPs}{Weakly Interacting Massive Particles}
\newabbreviation[category=common]{ROI}{ROI}{region of interest}
\newabbreviation[category=common]{pdf}{pdf}{probability density function}
\newabbreviation[category=common]{mixed}{mixed}{$\alpha$ + \textsuperscript{206}Pb}
\newabbreviation[category=common]{MOND}{MOND}{MOdified Newton Dynamics}
\newabbreviation[category=common]{MC}{MC}{Monte Carlo}
\newabbreviation[category=common]{CRESST}{CRESST}{Cryogenic Rare Event Search with Superconducting Thermometers}
\newabbreviation[category=common]{LNGS}{LNGS}{Laboratori Nazionali del Gran Sasso}
\newabbreviation[category=common]{SRIM}{SRIM}{Stopping and Range of Ions in Matter}
\newabbreviation[category=common]{TES}{TES}{Transition Edge Sensor}
\newabbreviation[category=common]{SOS}{SOS}{silicon-on-sapphire}
\newabbreviation[category=common]{SQUID}{SQUID}{Superconducting QUantum Interference Device}
\newabbreviation[category=common]{SIMS}{SIMS}{Secondary Ion Mass Spectrometry}
\newabbreviation[category=common]{TUM}{TUM}{Technische Universit\"at M\"unchen}
\newabbreviation[category=nameonly]{Geant4}{Geant4}{}
\newabbreviation[category=nameonly]{Root}{Root}{}
\newabbreviation[category=nameonly]{CADMesh}{CADMesh}{}
\newabbreviation[category=nameonly]{ImpCRESST}{ImpCRESST}{}
\newabbreviation[category=nameonly]{CresstDS}{CresstDS}{}
\newabbreviation[category=nameonly]{Bliss}{Bliss}{}
\newabbreviation[category=nameonly]{BAT}{BAT}{}
\newabbreviation[category=common]{RSM}{RSM}{Rough Surface Module}
\newabbreviation[category=nameonly]{cpp}{C++}{}
\newabbreviation[category=nameonly]{multiunion}{G4MultiUnion}{}
\newabbreviation[category=common]{RAM}{RAM}{Random-Access Memory}
\newabbreviation[category=nameonly]{LEICA}{LEICA DCM8}{}
\newabbreviation[category=common]{SNR}{SNR}{Screened Nuclear Recoil}
\newabbreviation[category=common]{Opt4}{EMOption4}{EMStandardPhysicsOption\_4}
\newabbreviation[category=nameonly]{SciPy}{SciPy}{}
\newabbreviation[category=nameonly]{Cuore}{CUORE}{}
\newabbreviation[category=nameonly]{Cosine}{COSINE-100}{}
\newabbreviation[category=nameonly]{Amore}{AMoRE}{}
\newabbreviation[category=common]{SurfaceLib}{SCoRe4}{Surface Contamination \& Roughness Effects for Geant4}
\begin{document}
\title[Article Title]{Geant4 based library "\gls{SurfaceLib}" for Surface Contamination and Roughness Effects simulations in rare event search experiments}
\author*[1,2]{\fnm{Christoph} \sur{Gr\"uner}}\email{christoph.gruener@oeaw.ac.at} 
\affil[1]{\orgdiv{Atominstitut}, \orgname{Technische Universit\"at Wien}, \orgaddress{\street{Stadionallee 2}, \city{Vienna}, \postcode{1020}, \country{Austria}}}
\affil[2]{\orgdiv{Marietta Blau Institute}, \orgname{\"Osterreichische Akademie der Wissenschaften}, \orgaddress{\city{Vienna}, \postcode{1010}, \country{Austria}}}
\abstract{
Surface simulations are important for accurately modeling particle interactions in experiments where background contributions from surface contaminants can significantly affect detector performance. 
In rare event searches, such as dark matter or neutrinoless double beta decay experiments, standard Geant4 simulations typically assume perfectly smooth surfaces, neglecting the microscopic roughness that exists in real materials.
This simplification can lead to inaccurate predictions of energy deposition.
To address this limitation, I developed \gls{SurfaceLib}, a Geant4-based library designed to simulate more realistic surface roughness based on experimentally measurable parameters.
The code allows users to generate patches of simplified rough surface geometries across a wide range of scales — from square millimeters to square meters — while maintaining computational efficiency.
\gls{SurfaceLib} is open source and can be easily integrated into existing Geant4 setups.
This work presents the structure, implementation, and example application of \gls{SurfaceLib},as well as its potential use in improving the accuracy of background modeling in rare event physics.
}
\keywords{Surface Roughness Simulation, Surface Contamination, Rare Event Search, Dark Matter, Neutrinoless Double Beta Decay, Geant4}
\maketitle
\section{Introduction}\label{sec:introduction}
\gls{Geant4} (GEometry ANd Tracking) is a widely used toolkit for simulating the propagation and interaction of particles, atoms, etc. with various geometries and materials \cite{geant4_1,geant4_2,geant4_3} based on Monte Carlo methods.
In the field of rare event research, such as in dark matter searches or neutrinoless double beta decay experiments, it is common to construct electromagnetic background models to better understand detector performance and background \cite{cresst_background,Alduino_2017}.
Therefore, the geometries and materials of the experimental setup are implemented in \gls{Geant4} and various sources that contribute to the background spectrum are added to the simulation.
This often includes atmospheric muons, ambient $\gamma$-rays, long-lived radiogenic impurities that are activated by cosmic rays.
This also involves bulk and surface contaminants such as $^{210}$Pb or $^{238}$U, found in the materials used for the experimental setup.
Surface contamination is of particular interest, since unlike bulk contamination—where nuclear decays produce a well-defined energy signature—the signal from surface contamination can be distorted or suppressed, as decay products may escape the detector, depositing only a fraction of their energy.\par
\gls{Geant4} offers a range of tools, settings and geometries to model a realistic setup.
The user can utilize different volumes like boxes, tubes, polyhedra, etc. combine or subtract them from one another to assemble a representation of the real-world geometry.
This results in complex geometric structures.
However, the surfaces of the geometries used are flat at the microscopic level. 
In contrast, actual surfaces exhibit a degree of roughness that is not accounted for in these simulations.
Due to this mismatch with reality, the effects of the surface structure can not be accounted for in the simulation.
This paper introduces the \gls{Geant4}-based library \gls{SurfaceLib}, designed to simulate surface roughness based on measured parameters and applicable to virtually any dimension of surface area from  a few \si{\milli\meter\squared} to \si{\meter\squared}, while maintaining computational efficiency.\par
This library is open source and published under the GPL-3.0 license, hosted on github \cite{score} and major versions are published on Zenodo \cite{gruner_2025_17648567}.
\gls{SurfaceLib} depends on several publicly available libraries and toolkits:
For simulations \textit{Geant4} (version 10.6.3) \cite{geant4_1, geant4_2, geant4_3};
for calculations \textit{Numpy}, \textit{SciPy} and \textit{Numba} \cite{harris2020array, 2020SciPy-NMeth, numba}; for data processing and representation \textit{Matplotlib}, \textit{Pandas}, \textit{PyYaml}, \textit{lxml} and \textit{trimesh} \cite{Hunter:2007, mckinney-proc-scipy-2010_pandas, pyyaml, behnel2005lxml, trimesh}; and for plotting of progress bars \textit{tqdm} \cite{casper_da_costa_luis_2024_14231923}.\par
This work is based on \gls{SurfaceLib} release version 1.0.0 \cite{gruner_2025_17648567} and is structured as follows:
An initial application of \gls{SurfaceLib} is discussed in \autoref{sec:backgroundModel}.
Multiple library code modules have been developed to address all aspects of simulating a rough surface, including the generation of rough surface geometry, scaling the geometry to represent nearly any surface size, and sampling vertices on and beneath the surface for contamination simulations.
These library modules are elaborated in \autoref{sec:library}.
In conclusion, \autoref{sec:conclusion} provides a future perspective.
An example of applying the library can be found in \autoref{sec:application}.\par
%
%
\section{Background Model}\label{sec:backgroundModel}
Background models in rare event search experiments are important to learn more about all sources contributing to the energy deposition spectrum and to either reduce their contribution or distinguish a genuine signal from them.
The development of a comprehensive and detailed background model that includes all important origins that contribute to the measured spectrum is a tedious task that involves multiple steps.
Especially detailed contributions from surface contaminations are hard to model based on the measured surface.\par
Here, the developed library \gls{SurfaceLib} adds the important detail of surface contributions to the spectrum.
The library allows for detailed simulation of particles that interact with a complex surface structure defined as a geometry in \gls{Geant4}.
This typically computer resource-demanding geometric representation of the surface involving millions of objects is feasible due to advancements within the library.\par
Incorporating these surface simulations into a background model, such as that developed by the \gls{CRESST} experiment \cite{cresst_background}, requires integrating \gls{SurfaceLib} into the experimental background model development workflow.
For the \gls{CRESST} experiment, \gls{SurfaceLib} was used to add surface roughness to the recreation of the experimental configuration within the \gls{Geant4} framework.
To create a background model the workflow includes the implementation of detectors, shielding, and various other components that have contaminants or that might contribute in other ways to the measured background.
Suitable \gls{Geant4} physics lists must be set for different interactions or physics such as the Coulomb interaction or decay physics.
The simulation domain incorporates multiple internal and external particle sources.
The aggregate energy deposited by each particle in the detector forms energy deposition templates for each particle type and source.
Subsequently, these templates are further processed to reflect the detector detection efficiency and then fitted to the experimental data.\par
The new library presented in this paper adds for the first time surface roughness to the experimental configuration within the \gls{Geant4} framework and allows for sources representing a contamination of these surfaces.
This opens up the possibility of accounting for more effects in the simulation, such as the surface roughness of the target, but also for the surrounding material, which is important for understanding the measured background in detail.\par
%
%
\section{Library}\label{sec:library}
\begin{figure}
    \centering
    \begin{overpic}[width=0.45\textwidth, right]{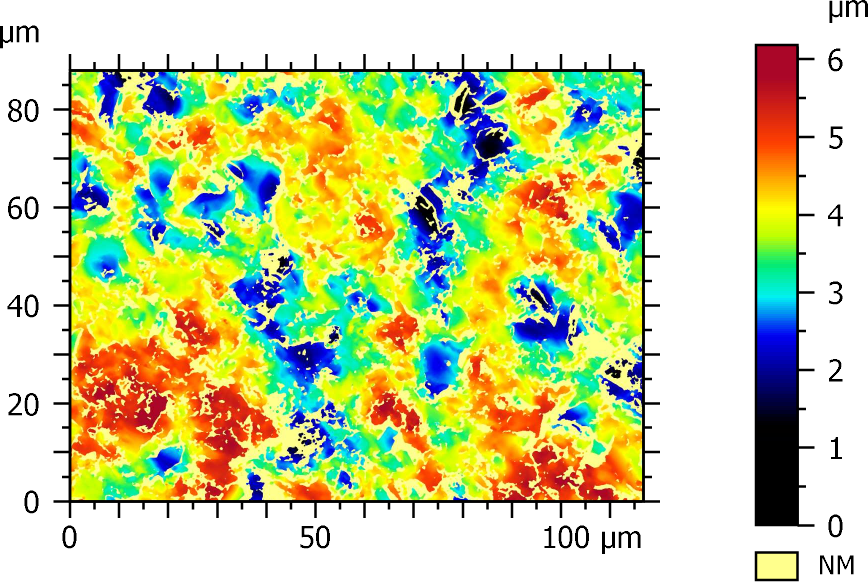}
        \phantomcaption
        \label{fig:TUM73}
        \put(85,-1){\color{white}\rule{30pt}{10pt}}
        \put(94,63){\color{white}\rule{20pt}{10pt}}
        \put(71,1.5){\color{white}\rule{20pt}{10pt}}
        \put(-0.1,61){\color{white}\rule{20pt}{10pt}}
        \put(99,15){\scriptsize\rotatebox{90}{profile height / \si{\micro\meter}}}
        \put(-6,18){\scriptsize\rotatebox{90}{Y-Range / \si{\micro\meter}}}
        \put(25,0){\scriptsize\rotatebox{0}{X-Range / \si{\micro\meter}}}
    \end{overpic}
    \vspace{2 mm}
    \caption{Surface profile of a diffused crystal surface. Structures in the \si{\micro\meter} scale are visible. Figure is taken from \cite{Gruener2024}}
    \label{fig:surface_profile}
\end{figure}
The surfaces of crystals used as target materials in rare event search experiment can be polished and therefore flat without nearly any structure or diffused with complex structures in the \si{\micro \meter} range as can be seen in \autoref{fig:surface_profile}.
Simulating a diffused contaminated surface of small scale (\si{\micro \meter}) structure can be tricky as even for a simplified representation of these structures by simple spikes, the number of needed spikes increases quadratically with surface area.
Having a surface area of only \SI{1}{\centi\meter\squared} and a spike base of the size of \SI{100}{\micro\meter\squared} already results in $10^6$ spikes to cover this area.
Covering slightly bigger areas or using spikes with a smaller base quickly exceeds the computers working memory.\par
Further, to simulate surface contamination the points to place the contaminants on and below the surface following some distribution depending on the depth below the surface must be defined.
Default \gls{Geant4} provides different particle generators to define such points, however, sampling on uneven surfaces cannot be done out of the box.
To address these problems, three different modules are implemented in the \textbf{\gls{SurfaceLib}: a) SurfaceRoughness, b) Portal, and c) ParticleGenerator}.
All modules combined allow one to perform detailed surface roughness contamination simulations.\par
\subsection{Module: Surface Roughness}
\label{subsec:Module:SurfaceRoughness}
\begin{figure}[ht]
  \centering
  \begin{subfigure}[t]{0.31\linewidth}
    \includegraphics[width=\linewidth]{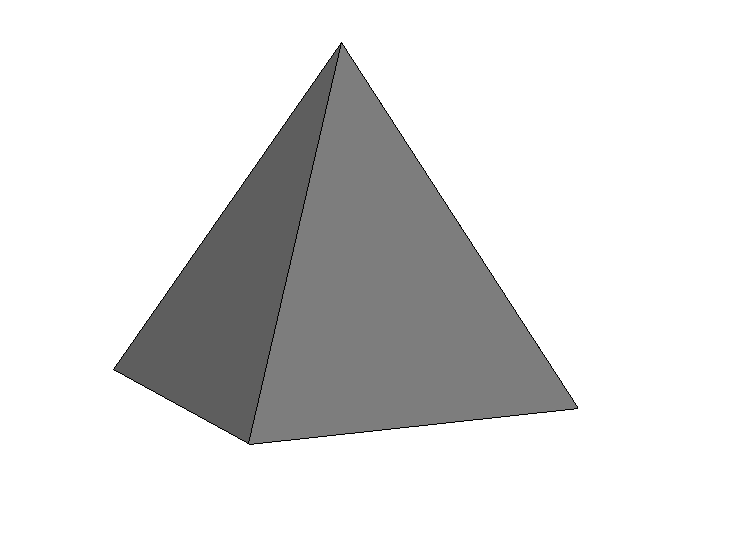}
    \caption{}
    \label{fig:spike_simple}
  \end{subfigure}\hfill
  \begin{subfigure}[t]{0.31\linewidth}
    \includegraphics[width=\linewidth]{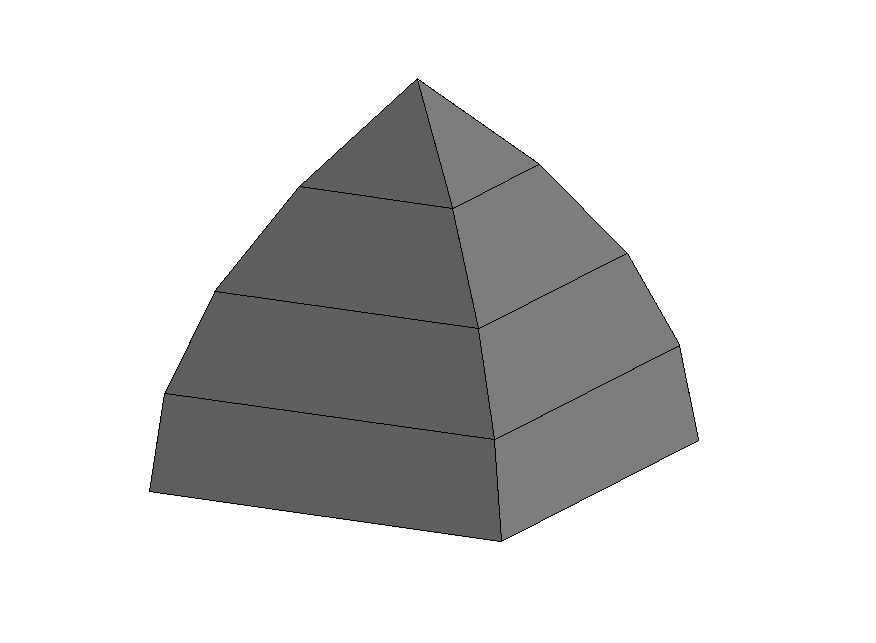}
    \caption{}
    \label{fig:spike_bump}
  \end{subfigure}\hfill
  \begin{subfigure}[t]{0.31\linewidth}
    \includegraphics[width=\linewidth]{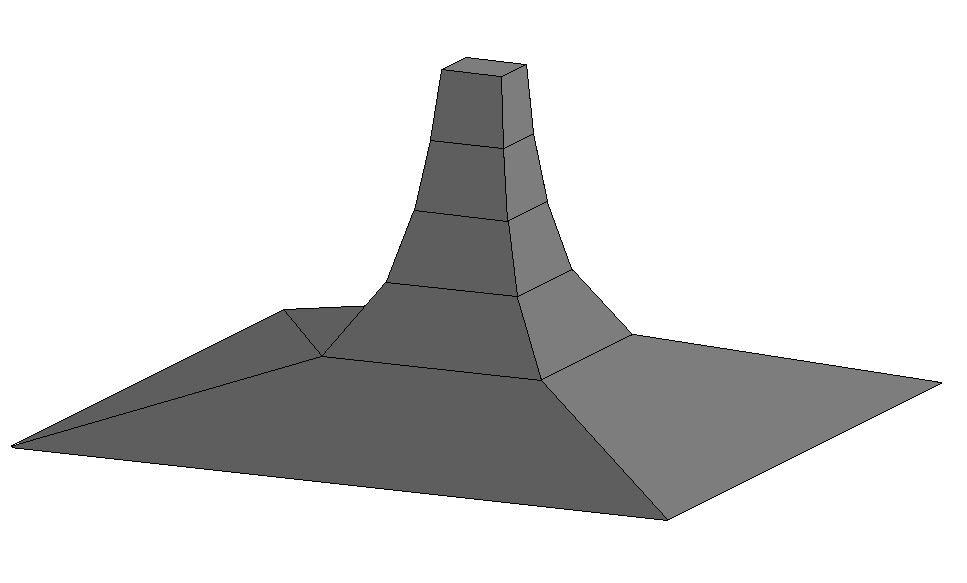}
    \caption{}
    \label{fig:spike_peak}
  \end{subfigure}
    \caption{Visualization of different shape implementations for spikes, which can be of any size that is allowed in \gls{Geant4}.: 
    (a) simplest form of a spike, basis is a \gls{Geant4} tetrahedron. 
    (b) multiple layers of \gls{Geant4} tetrahedrons form the spike, the outer surface approximates a squared function.
    (c) multiple layers of \gls{Geant4} tetrahedrons form the spike, the outer surface approximates $1/x$.}
  \label{fig:spike}
\end{figure}

\begin{figure}[ht]
    \centering
    \includegraphics[width=0.8\linewidth]{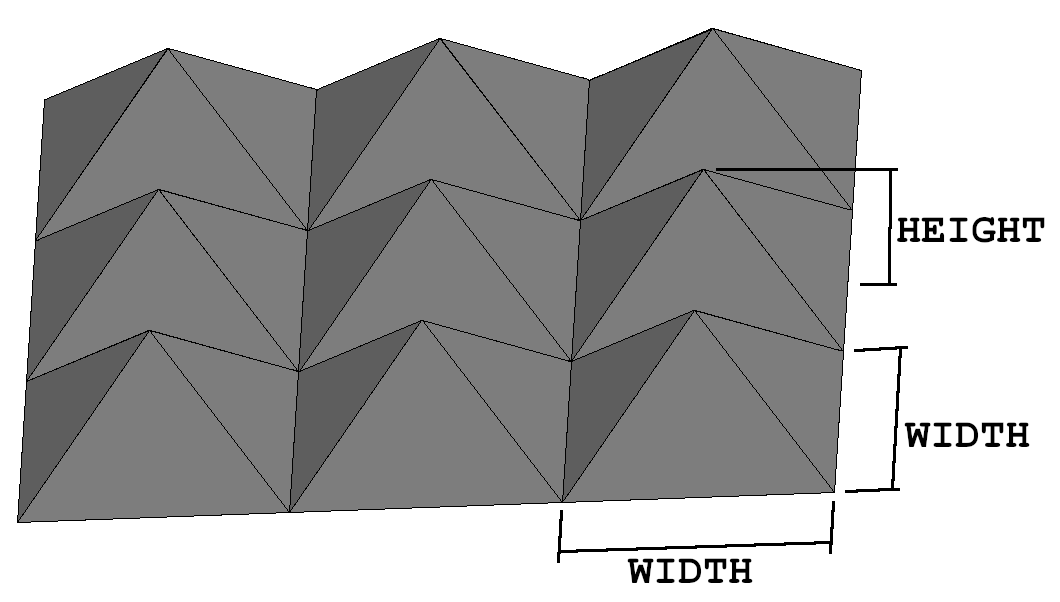}
    \caption{Visualization of 3x3 spikes placed at the surface of a target volume to simulate a patch of rough surface.
    The generated spikes can be of any size allowed in \gls{Geant4}.
    The size is controlled via setting a width and height for the spiked.}
    \label{fig:surface_batch}
\end{figure}

This module generates a rough surface patch represented by multiple spike-shaped structures (\autoref{fig:spike}) ordered in a grid like manner (\autoref{fig:surface_batch}).
A single structure can be either one physical volume or a combination of multiple physical volumes.
This allows for different-looking spike-like structures and a more complex surface.
However, all spikes in a batch still look alike.
The dimension of the basis of this structure must be the same for all spikes in a single patch of rough surface.\par
A surface patch is represented by a \gls{multiunion} and can contain up to $\sim1000 \times 1000$ spikes.
This number is limited by \gls{RAM} which arises due to the voxelization process for \gls{multiunion} (reference to voxelization).
It is an optimization algorithm that divides the simulation domain of \gls{multiunion} into small boxes (voxels) to improve the tracking of a simulated particle within the domain.
The number of boxes and, thus, the memory consumption increases with $\mathcal{O}(n^3)$ if the volume boundaries are not exactly on the same axis.
The maximum number of maximal splits in one dimension is hardcoded in Geant4 and set to $10^5$ and can result in $10^{15}$ boxes, which exceed the \gls{RAM} of a standard computer.
To change this number, the voxelization class was adapted.\par
The module also includes a calculator to compute the surface parameters \cite{iso25179}: Arithmetical Mean Height (Sa),
Root Mean Square Height (Sq), Skewness (Ssk) and Kurtosis (Sku).
\[
\renewcommand{\arraystretch}{1.5}
\begin{array}{r@{\hspace{0.4em}}l}
 Sa & =\frac{1}{A}\iint_A |z(x,y)| dA\\
 Sq & =\sqrt{\frac{1}{A}\iint_A z^2(x,y) dA} \\
Ssk & =\frac{1}{Sq^3}\left[\frac{1}{A}\iint_A z^3(x,y) dA\right]\\
Sku & =\frac{1}{Sq^4}\left[\frac{1}{A}\iint_A z^4(x,y) dA \right]
\end{array}
\]
\begin{figure}[ht]
  \centering
  \includegraphics[width=0.45\textwidth]{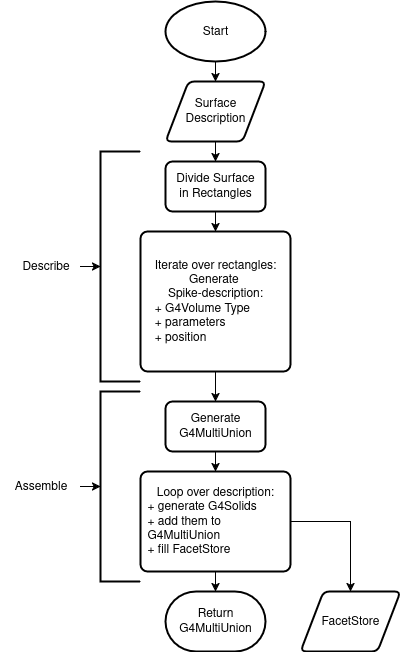}
  \caption{The flowchart shows the generation of a patch of rough surface, starting with passing the user defined surface description to the class.
  Next the chart is divided in two parts: 1) Describe represents the actions of the class \textit{Describer} which translates the user description into a list of needed \textit{G4Solids} inclusive their parameters and location.
  This list is passed to the \textit{Assembler} class which generates all the needed solids and adds them to a \textit{G4MultiUnion}.
  In parallel the \textit{FacetStore} is filled.}
  \label{fig:flowchart_surface_generation}
\end{figure}
The surface is described by a small bundle of parameters and must be set by the user:
\textbf{Spikeform} sets the general shape of the simulated spikes (see \autoref{fig:spike}).
\textbf{Width X/Y} describes the width of a single spike basis in the X and Y directions.
\textbf{Nr X/Y} describes the number of spikes in the X and Y directions.
\textbf{MeanHeight} sets the mean height of the spikes.
\textbf{Deviation} defines the one-$\sigma$ deviation of the height of the spikes from their mean height.
If set to zero, the mean height is the final height.
Some spike shapes require multiple layers to be formed, and the number of layers can be set with \textbf{Nr layer}.
All values can be either set via code during the generation of the detector or using a macro-file.\par
The whole generation of the patch of rough surface is controlled by the \textit{Generator} class.
This class handles the \textit{Describer} and \textit{Assembler} class (see \autoref{fig:flowchart_surface_generation}).
To generate \textit{G4MultiUnion}, the aforementioned parameters are passed on and interpreted by the \textit{Describer} class.
The class creates a list of all needed \textit{G4Solids} inclusive of their parameters (height, width, etc.) and relative position in the rough surface patch.
The \textit{Assembler} then creates the \textit{G4MultiUnion} and adds all listed solids.
In parallel, the FacetStore is filled with the correct facets that represent the exact surface.
Finally, the generator returns a handle to a \textit{G4MultiUnion}.
\subsection{Module: Portal}
\label{subsec:Module:Portal}
\begin{figure}[ht]
    \centering
    \vspace{-0.2cm}
    \begin{overpic}[width=1.\linewidth]{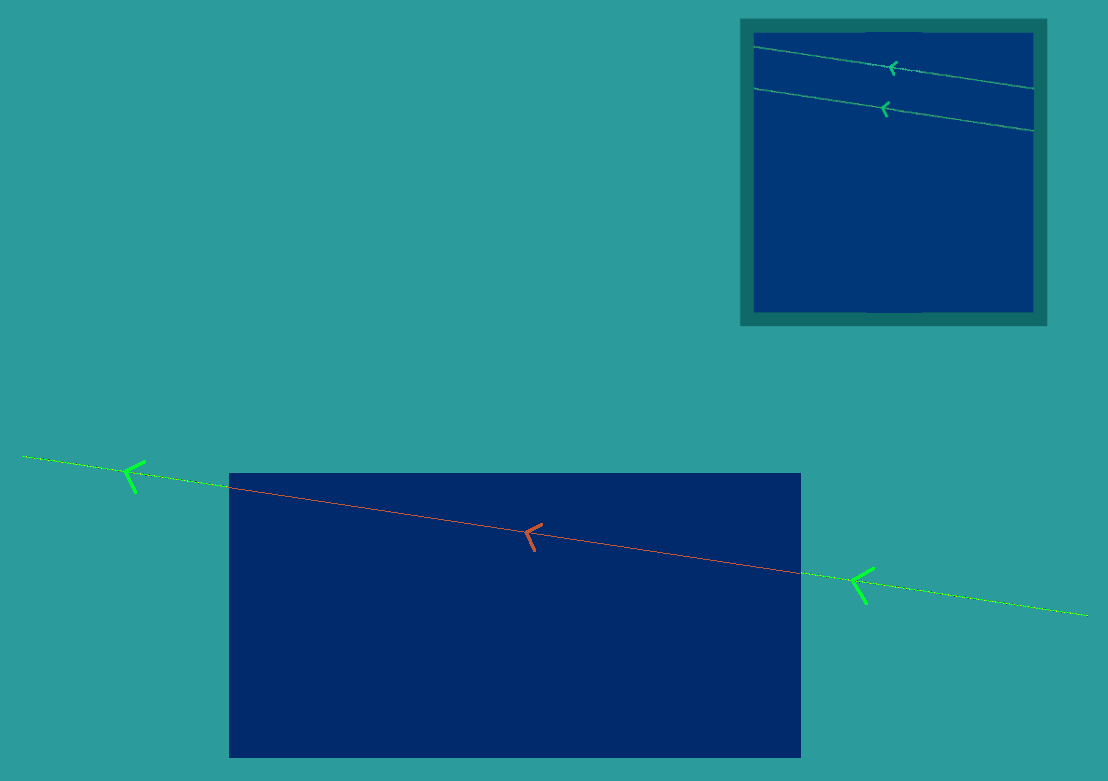}
        \put(57,55){\color{darkgray}\vector(1,0){9}}
        \put(32,39){\color{darkgray}\vector(1,-1){10}}
        \put(34,54){\textbf{Subworld}}
        \put(34,50){\textbf{+ Trigger}}
        \put(24,40){\textbf{Portal}}
        \put(73,14){\small\textbf{A$_1$}}
        \put(94.5,57){\small\textbf{A$_2$}}
        \put(61.5,61.5){\small\textbf{B$_1$}}
        \put(94.5,61){\small\textbf{B$_2$}}
        \put(61.5,65.5){\small\textbf{C$_1$}}
        \put(14,23){\small\textbf{C$_2$}}
        \put(47,1.8){\tikz \draw[dashed,red] (0,0)--(0,2);}
    \end{overpic}
    \caption{2D sketch of the function of the implemented Portal representing two subworlds. 
    The green dotted line is the trajectory of the particles entering the portal and the subworld on the right side and leaving them on the left.
    The orange dotted line is the imagined trajectory of the particle crossing the portal.
    The red dashed line divides the portal into two subworlds.\\
    The simulated particle undergoes the following steps: A) it enters the portal on the right side at A$_1$ and is ported to point A$_2$ of the subworld without a change in momentum. B) after crossing the subworld it leaves the volume at B$_1$ and enters the trigger. This activates the periodic teleportation and the particle is set to point B$_2$ without a change in momentum. C) after crossing the subworld for the second time it leaves the subworld's volume at C$_1$ and enters the trigger again. The particle is teleported to C$_2$ where it leaves the portal, without a change in momentum. 
    The dimension of the subworlds and the portal can be of any size allowed by \gls{Geant4}.
    The trigger volume ensures, a correct activation of particle teleportation.}
    \label{fig:portal}
\end{figure}

\begin{figure}[ht]
  \centering
  \includegraphics[width=0.45\textwidth]{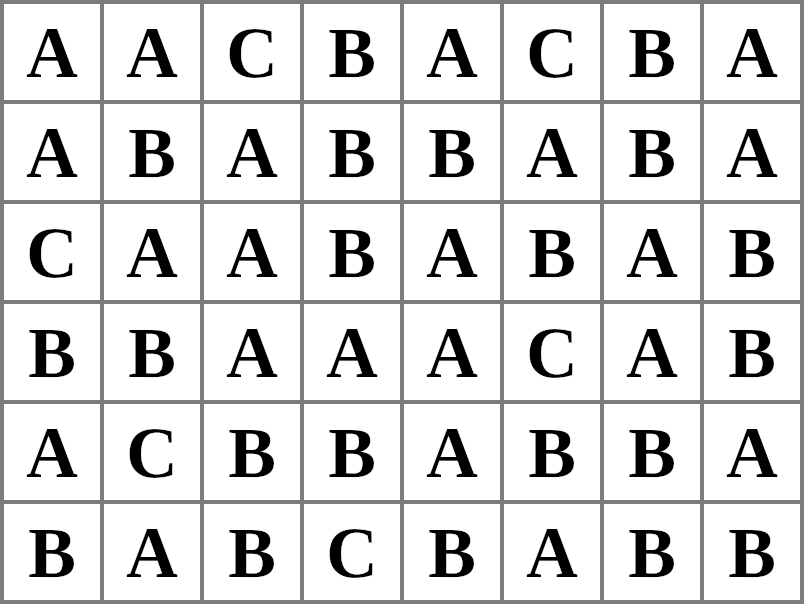}
  \caption{Subworlds can be combined in a grid to represent a more divers and bigger subworld.
  Thereby, already a small number of subworlds can be used to simulate a bigger domain.
  The grid can be filled manually or randomly with a user defined distribution.
  In this visualization three subworlds A, B and C are assembled in a $8 \times 6$ grid with set probabilities for A: 44\%, B: 44\%, C: 12\%.
  Because of the finite number of grid points, the final distribution of subworlds differs from the defined one.}
  \label{fig:subworld_grid}
\end{figure}

For the simulation of a diffuse / roughened surface, the surface roughness module \autoref{subsec:Module:SurfaceRoughness} simulates a patch of single spikes in the \si{\micro\meter} scale each.
To simulate at least a rough surface area of \SI{1}{\centi\meter\squared} this results in at least a million spikes.
Simulating bigger surface areas therefore exceeds computational resources such as random access memory, limiting the simulation domain to an order of \si{\milli\meter}.
In the following section, we introduce the portal-subworld structure, in which the portal volume is represented by a small number of subworlds that are reused to model the entire portal region (see \autoref{fig:portal}).
Placing the complex, computationally intensive geometry inside such a subworld allows the user to reuse it multiple times, without the high increase in needed resources.
Hereby, the portal volume acts as a placeholder and maps the covered volume to the corresponding single or multiple different subworlds.
For that, the portal volume contains a grid-like map that divides the volume into equally sized domains and assigns a subworld to it (see \autoref{fig:subworld_grid}).
A subworld can be assigned multiple times.
When a particle enters the portal volume, it is moved to the corresponding subworld on the basis of its entry position.
Inside the subworld the particle can interact with the geometries inside.
Upon leaving a subworld, the particle is either moved to the next corresponding subworld, re-enters the same subworld, or exits the portal volume at the corresponding position.
This is based on the position of the particle in the portal grid.
When a particle "moves" through the portal volume, it actually moves through the subwordls volume, but also does integer steps on the grid map when passing the subwordls boundaries.
This approach allows for the simulation of large and complex domains while requiring only a modest amount of memory.\par
The portal mechanism is controlled by the \textit{G4UserSteppingAction} class.
Each simulated particle step the mechanism checks if the particle must be moved.
This happens in two cases: the particle enters the portal or the trigger volume.
The trigger volume is part of the subworld and surrounds the subworlds simulation domain.
It guarantees for a fast and easy check that the particle left the subworld.\par
Unlike the portal volume, which is deliberately positioned within the simulation domain to enable particle interactions, the subworld must be located beyond the range of particles in the domain, since particles are intended to enter it exclusively through the movement mechanism governed by the \textit{G4UserSteppingAction}.\par
As setting up the portal and subwordls can be complex, portal-, subworld-, trigger-volumes must be created; a grid linking the portal and the subworlds must be filled; and the subworld-simulation domain placed inside its trigger.
To support the user, a helper class is provided.
The following parameters must be set: 
\textbf{Dx/Dy/DzSub} is the size of the subworld in x-, y- and z-direction,
\textbf{Dx/Dy/DzPortal} is the size of the portal in x-, y- and z-direction,
\textbf{Nx/Ny/Sub} sets the grid-size in x- and y-direction,
\textbf{nSubwords} sets the number of different subworlds,
\textbf{PortalName} and 
\textbf{SubworldName} define the portals and subworlds name,
with \textbf{MotherVolume} the mother volume is set,
\textbf{PortalPlacement} defines the global position of the portal,
\textbf{SubworldPlacement} is a vector of all global subworld positions,
\textbf{SubworldDensity} sets the density of the subworlds in the grid and
\textbf{SubworldMaterial} defines the material of the subworlds simulation domain.
\subsection{Module: Particle Generator/Shift}
\label{subsec:Module:ParticleGenerator}
With a portal-subworld setup and a rough surface in place, the sampling of the starting vertices with respect to the surface structure cannot be performed by \gls{Geant4}'s default particle generator.
Therefore, this library provides a custom particle generator that samples vertices uniformly distributed over the rough surface structure.
This works even in the case of a portal-subworld setup with different surface structures in different subworlds.\par
The uniform sampling is based on the facet store, introduced in \autoref{subsec:Module:SurfaceRoughness} and the portal grid from \autoref{subsec:Module:Portal}.
The facet store holds a representation of the exact outer surface of the surface roughness in the form of \textit{G4TriangularFacets}.
The sampling of uniformly distributed points is based on the surface area and executed in three steps: 1) uniform sampling of a subworld in the portal grid based on the total surface area in the corresponding subworld; 2) sampling of \textit{G4TriangularFacet} in the corresponding facet store based on the facets area; 3) sampling of point on the facet using \gls{Geant4}s sampling mechanism.\par
The particle generator samples points exactly at the surface, however, to start particle simulations in the vicinity below the surface, the sampled vertices have to be shifted.
This is done by the shift module, which takes a sampled vertex and moves it in a certain direction.
The direction and size of the shift can be defined by the user.
However, in the case of surface contamination simulation, the direction is perpendicular to the surface at which the vertex is sampled, pointing below the surface.\par
\begin{figure}[ht]
    \centering
    \begin{overpic}[width=1.\linewidth]{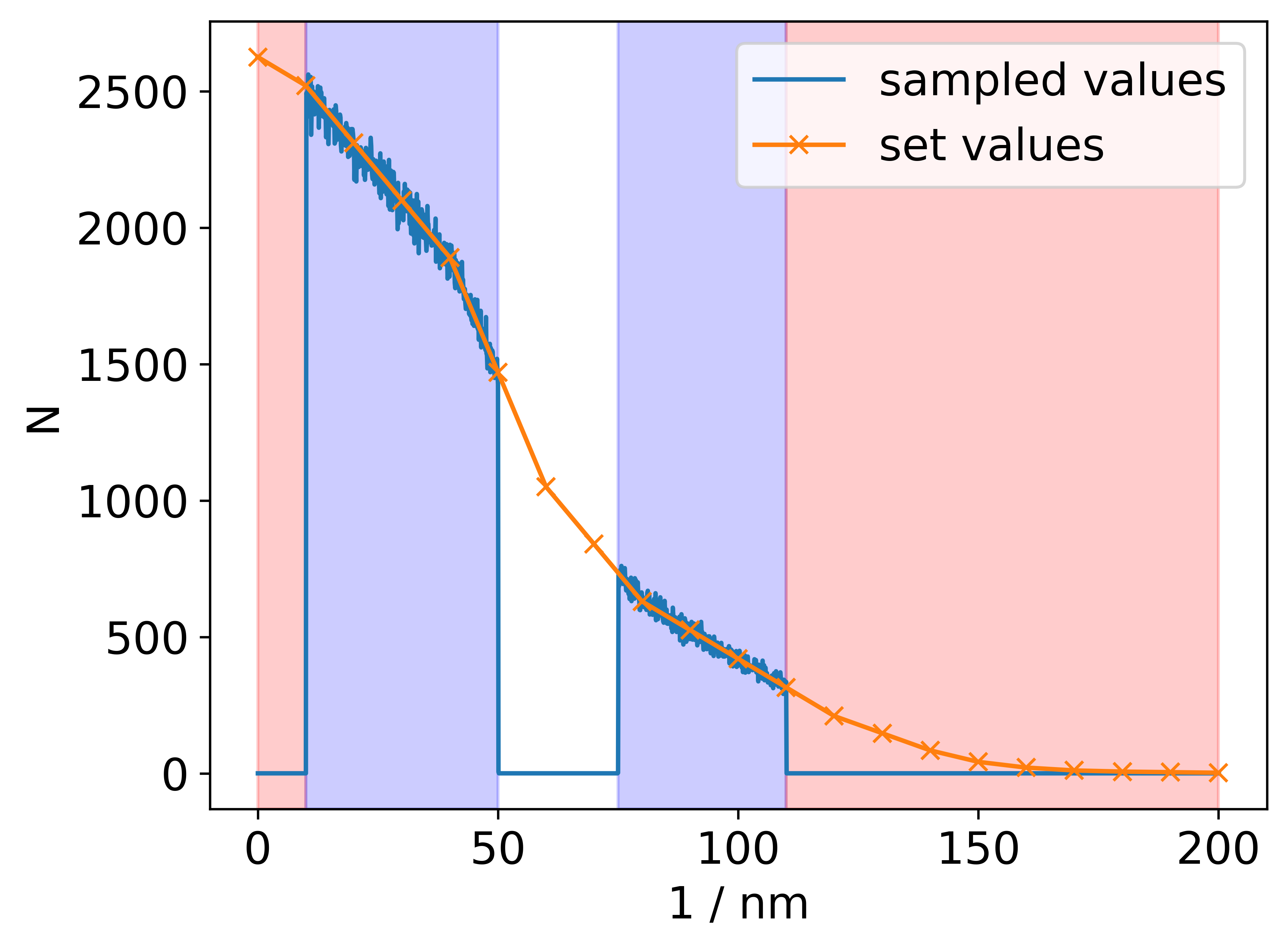}
        \put(19.5,20){\rotatebox{30}{\small\textbf{Volume A}}}
        \put(64.5,20){\rotatebox{30}{\small\textbf{Volume B}}}
        \put(38.5,22){\rotatebox{30}{\small\textbf{Gap}}}
    \end{overpic}
    \caption{Example of sampled values for a shift (blue) based on a defined distribution (orange, the x marker are the set values in the distribution file, the line represents the linear interpolation between them) in the range of \SI{}{\nano\meter}.
    In the example, two \gls{Geant4} volumes (marked as red and blue areas) are placed in line with a gap (50 to \SI{75}{\nano\meter}) between them. Although the distribution is non zero in the areas marked as red (0 to \SI{10}{\nano\meter}, 110 to \SI{200}{\nano\meter}) and white (50 to \SI{75}{\nano\meter}) no points are sampled in this ranges due to special options set for shifting:
    a) a minimal shift of \SI{10}{\nano\meter} is set, therefore no values are sampled in the first red marked area.
    b) a maximal shift of \SI{110}{\nano\meter} is set, therefore no values are sampled in the second red marked area.
    c) the shift is confined to the volumes material but the white area is a gap between the two volumes, therefore not points are sampled from (50 to \SI{75}{\nano\meter}).
    }
    \label{fig:shift_example}
\end{figure}

A sampling function can be passed to the module in the form of a histogram-like list of step in \si{\nano\meter} and number of expected vertices at this step.
In addition, sampling can be restricted to an area by setting a minimal and maximal shift or by restricting to a material.
The sampler linearly interpolates the missing values in between.
In \autoref{fig:shift_example} the actual sampling can be seen in comparison to the set values including a minimal (\SI{10}{\nano\meter}) and maximal (\SI{110}{\nano\meter}) shift and restriction to the volumes material.
\subsection{All modules combined}
All single modules complete a certain task, such as reusing a small simulation domain again to represent a larger one, creating a rough surface structure, or sampling and shifting staring vertices for simulation.
To perform a full surface contamination simulation of a rough surface, all of these modules must be linked together.
Extending the rough surface patch is rather simple as it must only be placed within the reused subworld.
In addition, when generating the rough surface patch, the module also creates the surface store.
This must be placed in the position of the rough surface patch.
An extra positioning is necessary, as the generated \textit{G4MultiUnion} representing the rough surface patch is a \textit{G4Solid} and does not have a final position.
It must be placed as a physical volume in the simulation domain.
Further, the facet store must be linked to the subworld just by passing the pointer of the facet store.
Finally, the name of the portal must be added to the sampler.
For the simplification of the setup process, helper classes are provided.
(For an example setup, see \autoref{sec:application}.)
\subsection{Extension Module: Complex Surface Generation}
\begin{figure}
    \centering
    \includegraphics[width=0.45\textwidth]{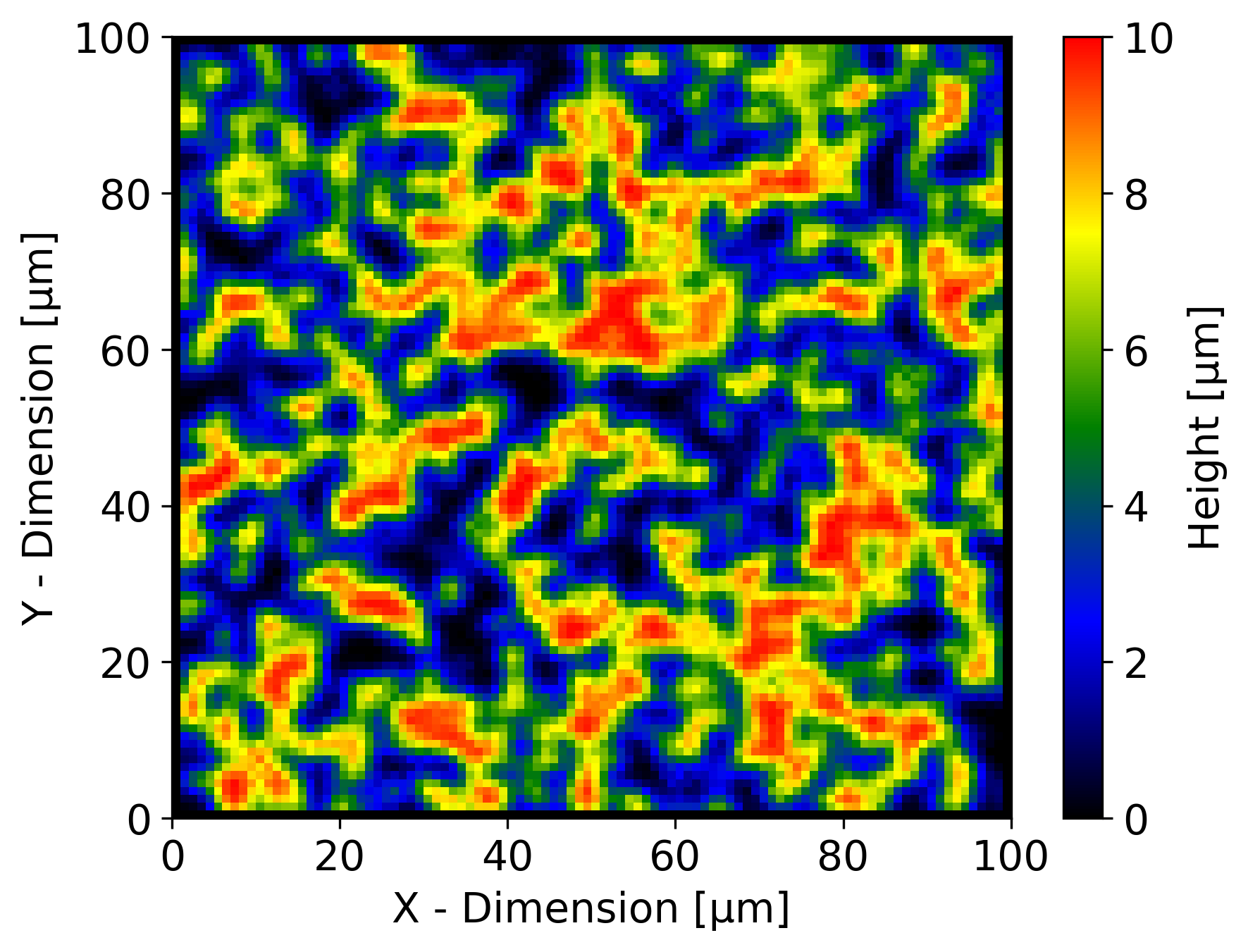}
    \caption{Example of surface profile generated with \gls{SurfaceLib}. Structures in the \si{\micro\meter} scale are visible.}
    \label{fig:generated_surface_profile}
\end{figure}
The introduced modules enable the simulation of a simplified patch of a rough surface over a large area using spikes.
However, the portal module is versatile and can be used with any geometry inside.
Therefore, more complex surfaces can be generated and utilized.
The additional Python tool \textit{SurfaceGenerator.py}, which is a part of \gls{SurfaceLib}, was developed for this purpose.
Using this tool, surface profiles (see \autoref{fig:generated_surface_profile}) can be generated and exported to a \textit{GDML} file \cite{gdml}.
The tool provides different routines to generate more complex surfaces and can be used via the command line.
In this file, the surface is stored as a \textit{G4Tessellated} object, which can be loaded into \gls{Geant4} using the class \textit{LogicalSurface}.
Although \textit{GDML} files can be loaded without the provided class, its usage is highly recommended, as the class is also used to sample evenly distributed points from the surface later in the \gls{Geant4} simulation process.\par
However, in comparison to the simpler surface representation using spikes and a portal, the use of \textit{G4Tessellated} objects increases computation time and is currently about two magnitudes slower.
An example of an application is shipped with the library.
%
%
\section{Conclusion}\label{sec:conclusion}
This paper presents the developed Geant4-based C++ library \gls{SurfaceLib} to add surface roughness in combination with surface contamination to simulations.
This allows, for the first time, the simulation of surface roughness based on real measured surface parameters.
This is especially important for rare event research experiments such as \gls{CRESST} \cite{cresst_background} or \gls{Cuore} \cite{Alduino_2017}, as the effects of surface contamination have an impact on the measured electromagnetic background spectrum.
Therefore, a detailed background simulation is of the utmost importance.
The library was developed and tested on the basis of the \gls{CRESST} experiment \cite{CRESST:2024chq,Gruener2024}, and I was further able to reproduce some results of the \gls{Cuore} experiment. For more information, see \cite{Gruener2024}.\par
The library \gls{SurfaceLib} comes with three C++ code modules: to simulate a small patch of rough surface; extend the patch of rough surface to a large surface area of your choice; and a particle generator to sample starting vertices on the patch of rough surface and shift them below the surface.
All these modules combined now allow for simulations of rough and contaminated surfaces.

\begin{figure}[ht]
    \centering
    \includegraphics[width=1\linewidth]{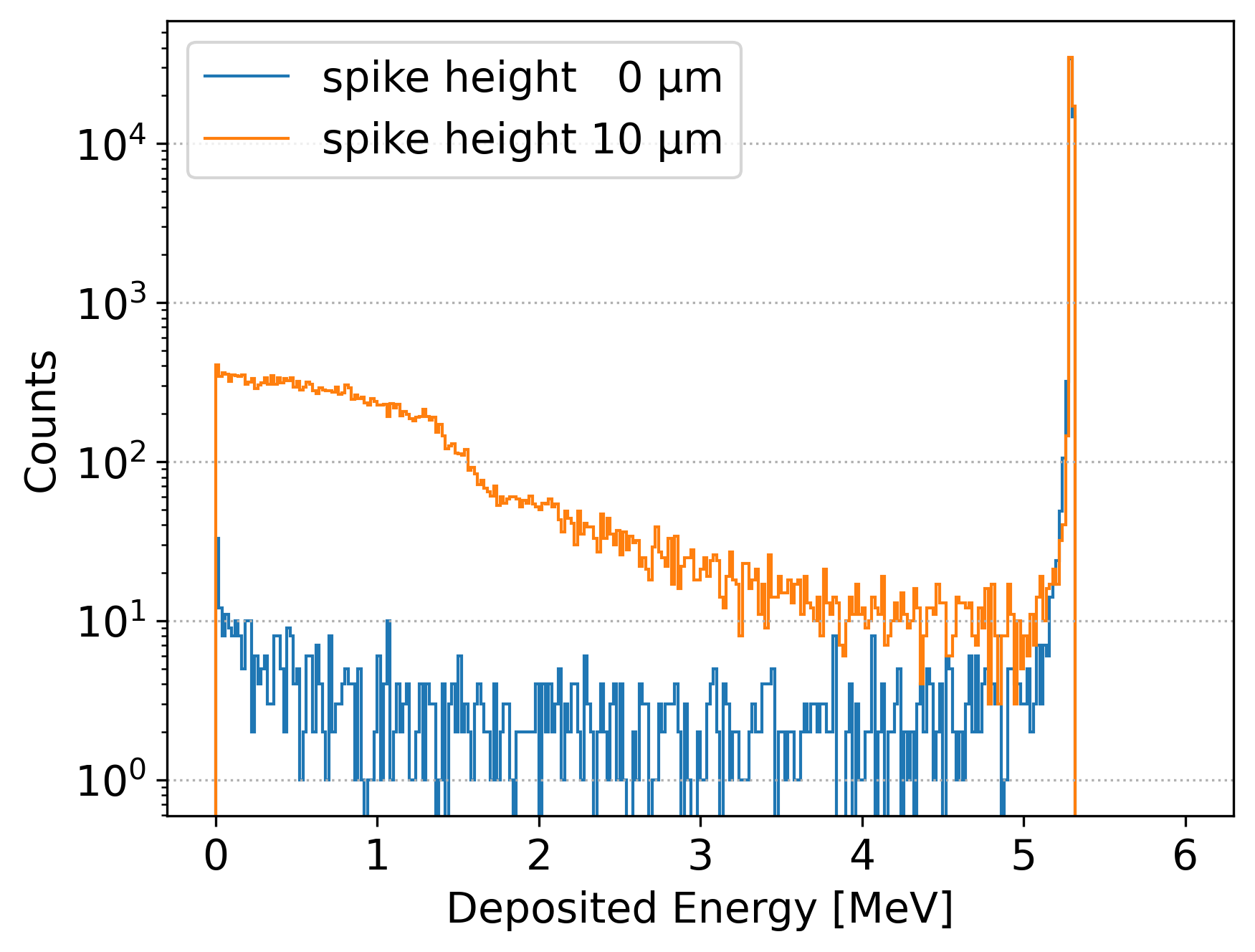}
    \caption{Energy deposition spectrum of $\alpha$-particles with an energy of \SI{5.3}{\mega\electronvolt} placed at a 3x\SI{3}{\centi\meter\squared} surface without and with spikes of height and width of \SI{10}{\micro\meter}.}
    \label{fig:deposited_energy_example}
\end{figure}

A simulation result is provided in \autoref{fig:deposited_energy_example} where $\alpha$-particles with an energy of \SI{5.3}{\mega\electronvolt} are distributed uniformly on a flat or rough 3x\SI{3}{\centi\meter\squared} surface of a silicon crystal.
This energy corresponds to the typical energy of $\alpha$-particles emitted during the decay of Po$^{210}$ in the Pb$^{210}$ decay chain, a potential surface contaminant \cite{Gruener2024}.
The rough surface is characterized by spikes measuring \SI{10}{\micro\meter} in both height and width.
The effect of a roughened surface is clearly visible for energies below \SI{4}{\mega\electronvolt} by a higher number of counts. 
\par
An additional Python code module provided with a command line interface can be used to generate a more complex surface and export it to a \textit{GDML} file.
This is supported by further Geant4-based C++ classes to correctly load the surface volume from the \textit{GDML} file, place it correctly in simulation and get access to uniform point sampling from its surface.
Despite offering more detailed surface simulations, this newly developed module demands significant computational resources, resulting in extended runtimes.\par
The open source library \gls{SurfaceLib} provides a foundation for surface roughness and contamination simulations in \gls{Geant4} and should pave the way toward a better understanding of the effects of microscopic surface structures in rare event research experiments.\par

Following this publication, the library will be maintained and further extended to improve simulation speed using \textit{G4Tessellated} surface objects, as well as to provide additional options for reconstructing more realistic surfaces.\par
Code examples and additional usage instructions are shipped with the library.\par
Interested users are encouraged to participate in testing, giving feedback, reporting bugs, or contributing to the continued development of the library \cite{score}.
%
%
\bmhead{Acknowledgements}
I would like to thank the CRESST collaboration for the many discussions and valuable input. 
In particular, I would like to acknowledge the simulation group of CRESST — Valentyna Mokina, Samir Banik, Holger Kluck, Jens Burkhart and Robert Breier — for their contributions. 
I am especially grateful to Valentyna Mokina for her continuous feedback on the progress of this paper.

This work has been funded in part by the Austrian Science Fund (FWF) by \url{http://dx.doi.org/10.55776/I5420}.

\appendix
%
%
\section{Example}\label{sec:application}
For setting up a simulation using \gls{SurfaceLib} multiple steps must be completed.
In the following section, an example is presented and important details are discussed.
The goal is to establish a rough surface patch of silicon, an area of \SI{10}{\centi\meter} x \SI{10}{\centi\meter} and a roughness in the range of \si{\micro\meter}.
In a next step, we introduce $^{210}$Po as contamination on the surface.
Setting up this simulation involves: the preparation and linking of a portal and subworld; the activation of the teleportation mechanism; the definition of a rough surface and its placement in the subworld; and the setup of the particle generator.
The most important details of each step are discussed.
For a description of the named modules, see section \autoref{sec:library}.
More examples and further explanation can be found with the code \cite{score,gruner_2025_17648567}.\par
\subsection{Setup of the Simulation}\label{subsec:Setup}
In the following, the most important steps in preparing the simulation are presented.
It is assumed that the user has some familiarity with the \gls{Geant4} simulation framework, particularly with \textit{the G4VUser...} classes, such as \textit{G4VUserDetectorConstruction} or \textit{G4VUserPrimaryGeneratorAction}.
\subsubsection{Portal - Subworld}\label{subsubsec:Portal-Subworld}
The key element for a surface simulation with macroscopic areas is the portal-subworld module.
To simplify its setup, a helper class \textit{Surface::MultiportalHelper} is provided, which first gathers all the needed information, such as the size of the portal, size of the subworld, number of different subworlds and their frequency, materials, etc.
After providing the information, the helper generates a portal-subworld pair (\autoref{lst:portal-subworld}).\newline
To activate the portal mechanism, the control class must be added to the \textit{G4UserSteppingAction} and activated by calling the \textit{DoStep(...)} function. (\autoref{lst:activation}).
\begin{lstlisting}[language=C++, caption={Setup of Portal-Subworld pair. Class: G4VUserDetectorConstruction}, label={lst:portal-subworld}]
#include "MultipleSubworld.hh"
#include "MultiportalHelper.hh"

G4VPhysicalVolume *DetectorConstruction::Construct() {
 ...

  Surface::MultiportalHelper helper("Helper", 5);
  
  //definde portal
  helper.SetPortalName("Portal");
  helper.SetDxPortal(10 * cm);
  helper.SetDyPortal(10 * cm);
  helper.SetDzPortal(0.5 * cm);
  helper.SetPortalPlacement(trafoPortal);

  //set size of subworld
  helper.SetDxSub(0.5 * mm);
  helper.SetDySub(0.5 * mm);
  helper.SetDzSub(0.5 * cm);

  //set number of subwordls (must match with subworld- and portal size)
  helper.SetNxSub(200);
  helper.SetNySub(200);

  //set number of different subworlds and material
  helper.SetNDifferentSubworlds(3);
  helper.SetSubworldMaterial(subworldMaterial);

  //place subworlds
  helper.AddSubworldPlacement(trafoA); 
  helper.AddSubworldPlacement(trafoB);
  helper.AddSubworldPlacement(trafoC);

  //set density of subworlds
  helper.AddSubworldDensity(0.3);
  helper.AddSubworldDensity(0.5);
  helper.AddSubworldDensity(0.2);

  //set mother volume for portal-subworld setup
  helper.SetMotherVolume(logicWorld);

  //generate setup
  helper.Generate();
  ...
  return physWorld;
}


\end{lstlisting}
\begin{lstlisting}[language=C++,caption={Activation of Portal-Subworld, File: SteppingAction.cc}, label={lst:activation}]
#include "PortalControl.hh"

void SteppingAction::UserSteppingAction(const G4Step *step){
    fPortalControl.DoStep(step);
}
\end{lstlisting}
\subsubsection{Surface Structure}\label{subsubsec:SurfaceStructure}
The generation of a rough surface can be accomplished by using the \textit{Surface::RoughnessHelper} class (\autoref{lst:surface}).
Based on a description (number of spikes, their size and shape, etc.) that can be done in code or via a macro file (\autoref{lst:surface_macro}), this class returns a \textit{G4LogicalVolume} that represents a rough surface patch and must be handled accordingly.

\begin{lstlisting}[language=C++,caption={Generation of surface based on macro file}, label={lst:surface}]
G4VPhysicalVolume *DetectorConstruction::Construct() {
...
f_surface_helper.Generate();
G4LogicalVolume *logical_roughness = f_surface_helper.LogicRoughness();
...
}
\end{lstlisting}
In our setup, we introduce macro files similar to those used in \gls{Geant4}.
These macro files act as control scripts for key simulation options, such as enabling or adjusting surface roughness generation, or configuring the portal subworld setup.
The main benefit is flexibility: Parameters can be changed directly in the macro file without modifying or recompiling the source code, making testing faster and easier.\par
\begin{lstlisting}[caption={Macro commands for surface generation}, label={lst:surface_macro}]
/Surface/RoughnessHelper/setVerbose 2

/Surface/RoughnessHelper/setBasisDx 0.5 mm
/Surface/RoughnessHelper/setBasisDy 0.5 mm
/Surface/RoughnessHelper/setBasisDz 0.25 mm

/Surface/RoughnessHelper/setSpikeDx 5 um
/Surface/RoughnessHelper/setSpikeDy 5 um
/Surface/RoughnessHelper/setSpikeMeanHeight 5 um
/Surface/RoughnessHelper/setSpikeDevHeight 1 mm
/Surface/RoughnessHelper/setSpikeform StandardPyramid

/Surface/RoughnessHelper/setSpikesNx 100
/Surface/RoughnessHelper/setSpikesNy 100

/Surface/RoughnessHelper/setMaterial G4_Si

/Surface/RoughnessHelper/setBoundaryNx 1000
/Surface/RoughnessHelper/setBoundaryNy 1000
/Surface/RoughnessHelper/setBoundaryNz 1000
\end{lstlisting}
\subsubsection{Particle Generator}\label{subsubsec:ParticleGenerator}
The particle generator can evenly distribute points on the rough surface with respect to its placement in a portal subworld.
For that, the generated rough surface must be linked to the portal.
This can be done using the \textit{Surface::MultiportalHelper} and \textit{Surface::RoughnessHelper} classes \autoref{lst:particle_generator}.
Afterwards, the \textit{Surface::MultiSubworldSampler} class can be used for sampling points in \textit{G4VUserPrimaryGeneratorAction}.
\begin{lstlisting}[language=C++,caption={Setup of particle generator}, label={lst:particle_generator}]
G4VPhysicalVolume *DetectorConstruction::Construct() {
...
f_surface_helper.Generate();

G4Transform3D trafo_surface{ G4RotationMatrix(), G4ThreeVector(0., 0., +f_surface_helper.GetBasisHeight() * 2 - f_portal_helper.GetPortalDz())};

f_portal_helper.AddRoughness( f_surface_helper.LogicRoughness(), trafo_surface, f_surface_helper.FacetStore());

f_portal_helper.Generate();
...
}
\end{lstlisting}
%
%
\bibliography{sn-bibliography}
\end{document}